\documentclass[letterpaper]{article} 
\usepackage{aaai24}  
\usepackage{times}  
\usepackage{helvet}  
\usepackage{courier}  
\usepackage[hyphens]{url}  
\usepackage{graphicx} 
\usepackage{natbib}  
\usepackage{caption} 
\usepackage{algorithm}
\usepackage{algorithmic}
\usepackage{amsfonts} 
\usepackage{amsmath}
\usepackage{booktabs}
\usepackage{siunitx}
\usepackage{amssymb}
\usepackage{subcaption} 

\usepackage{newfloat}
\usepackage{listings}
\DeclareCaptionStyle{ruled}{labelfont=normalfont,labelsep=colon,strut=off} 
\floatstyle{ruled}
\newfloat{listing}{tb}{lst}{}
\floatname{listing}{Listing}
\title{Phoneme Hallucinator: One-shot Voice Conversion via Set Expansion}
\author{
    Siyuan Shan, Yang Li, Amartya Banerjee, Junier B. Oliva
}
\affiliations{
    Department of Computer Science, University of North Carolina at Chapel Hill\\

    Chapel Hill, North Carolina, USA\\
    \{siyuanshan, yangli95, amartya1, joliva\}@cs.unc.edu
}

\usepackage{bibentry}

\begin{document}

\maketitle

\begin{abstract}
Voice conversion (VC) aims at altering a person's voice to make it sound similar to the voice of another person while preserving linguistic content. Existing methods suffer from a dilemma between content intelligibility and speaker similarity; i.e., methods with higher intelligibility usually have a lower speaker similarity, while methods with higher speaker similarity usually require plenty of target speaker voice data to achieve high intelligibility. In this work, we propose a novel method \textit{Phoneme Hallucinator} that achieves the best of both worlds. Phoneme Hallucinator is a one-shot VC model; it adopts a novel model to hallucinate diversified and high-fidelity target speaker phonemes based just on a short target speaker voice (e.g. 3 seconds). The hallucinated phonemes are then exploited to perform neighbor-based voice conversion. Our model is a text-free, any-to-any VC model that requires no text annotations and supports conversion to any unseen speaker. Objective and subjective evaluations show that \textit{Phoneme Hallucinator} outperforms existing VC methods for both intelligibility and speaker similarity. 
\end{abstract}

\section{Introduction}
\label{sec:intro}
Voice conversion (VC) refers to the process of altering the acoustic characteristics of a person's voice (source) to another person's voice (target) while preserving linguistic content \cite{mohammadi2017overview}. It has a wide range of applications such as the creation of personalized voices for those with speech disabilities, vocal identity protection, and entertainment purposes on short-video platforms. \textit{Intelligibility} and \textit{speaker similarity} are two important criteria to evaluate a VC model, the former measures the degree to which spoken words or utterances in the converted speech are clear and understandable to listeners, while the latter reflects the similarity between the converted speech and the target speech. A good VC model should achieve high intelligibility and produce speech characteristics similar to target speakers simultaneously.  

\begin{figure}[t!]
\centering
\includegraphics[width=\linewidth]{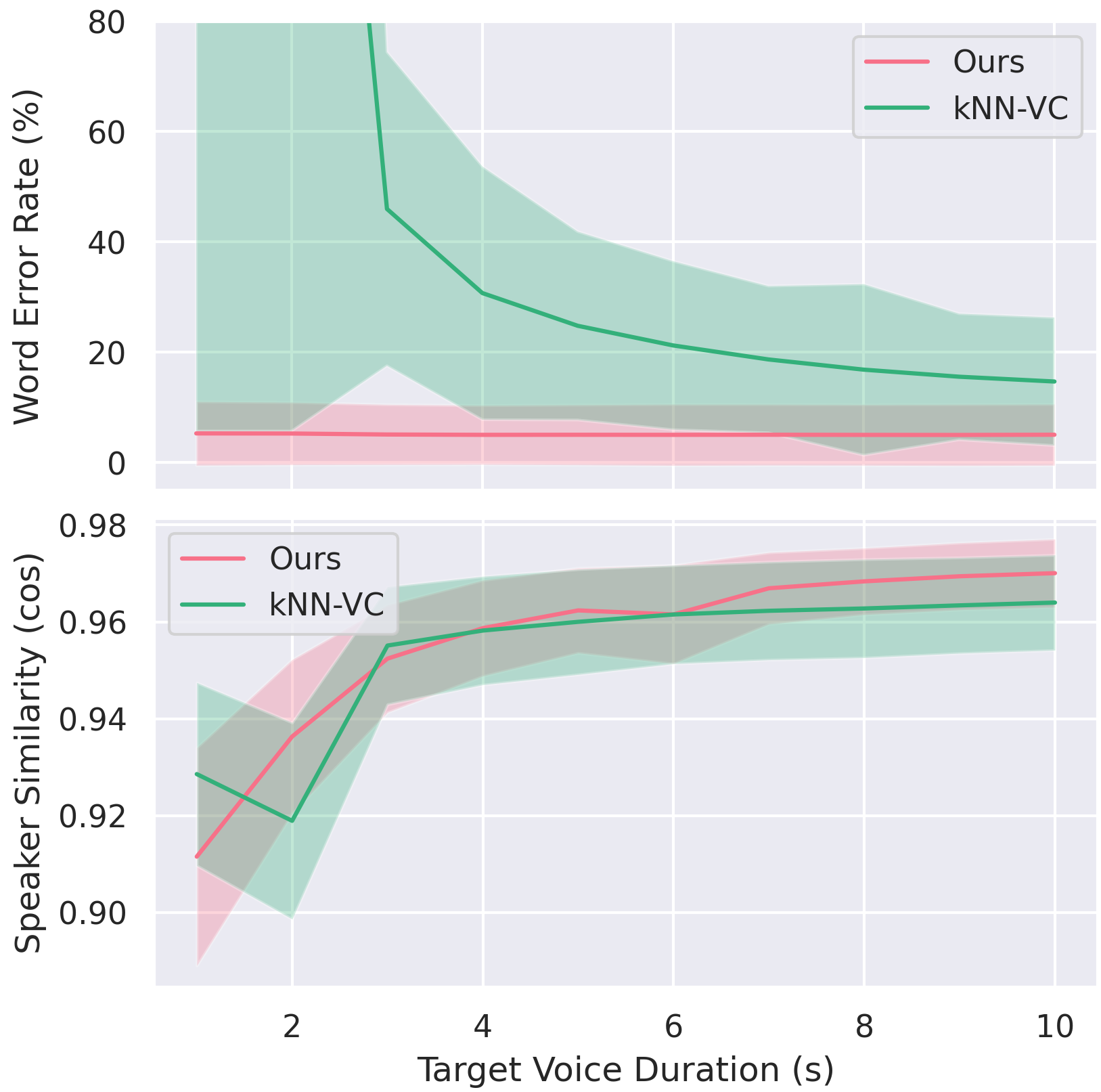} 
\caption{Comparison between kNN-VC (SOTA method) and our proposed \textit{Phoneme Hallucinator} on the LibriSpeech \textit{test-clean} split with varying target voice duration. Word Error Rate (WER$\downarrow$) and Speaker Similarity ($\uparrow$) respectively measure intelligibility and speaker similarity. Shadowed areas indicate standard deviations of WER and Speaker Similarity computed on the \textit{test-clean} split. }
\label{fig1}
\end{figure}

There have been significant advances for VC \cite{li2023freevc,qian2019autovc,chou2019one,casanova2022yourtts} recently due to the development of novel generative models such as VAE, GAN, flow models, and diffusion models. These methods adopt speaker embedding models \cite{snyder2018x} to extract target speaker characteristics and rely on designs such as information bottlenecks \cite{qian2019autovc}, normalization methods \cite{chou2019one}, or data augmentation \cite{li2023freevc} to convert speech.  However, they still struggle with speaker similarity as shown in a recent work kNN-VC \cite{baas2023voice}. Instead of using a speaker embedding model and complicated designs, kNN-VC introduces a simple method. First, it extracts frame-wise self-supervised representations for the source and the target speech respectively. As the representations contain both speaker characteristics and phoneme information, conversion can be performed by directly replacing each frame of the source representation with its nearest neighbors in the target representation set. Finally, the audio is synthesized by a vocoder that takes the converted representation as input. Compared to speaker embedding VC methods, kNN-VC achieves better speaker similarity because the output is constructed exclusively from target speech.

However, we find the intelligibility of kNN-VC deteriorates severely when the target voice is short as shown by its high Word Error Rate in Fig. \ref{fig1}. This problem may negatively impact its real-world application where users usually can only provide several seconds of target speech. The cause of this problem is straightforward: for short target voices, only a small number of self-supervised representations can be extracted. As a result, some key neighbors with necessary phonemes are missing for high-quality neighbor replacement.

In this work, we propose the \textit{Phoneme Hallucinator}, a conditional generative model that can hallucinate target representation features conditioned on a small target representation set. With more target features, the target neighbor candidates set is effectively expanded and the intelligibility of the converted speech is improved without sacrificing the speaker similarity as indicated in Fig. \ref{fig1} and our extensive experiments. Our contributions are as follows:
\begin{itemize}
    \item We propose \textit{Phoneme Hallucinator}, a principled generative model inspired from the perspective of probabilistic set modeling. 
    \item Our novel generative model is seamlessly integrated into a neighbor-based VC pipeline to ``hallucinate'' speech representations to achieve one-shot VC.
    \item Extensive experiments demonstrate that \textit{Phoneme Hallucinator} achieves SOTA performance regarding both intelligibility and speaker similarity on one-shot VC tasks where only a 3-second of target speaker speech is given.
    \item Codes and audio samples are available at \url{https://phonemehallucinator.github.io/}.
\end{itemize}

\section{Related Work}
FreeVC \cite{li2023freevc} and kNN-VC \cite{baas2023voice} are two representative SOTA text-free, any-to-any VC methods, which means they don't require any text annotations and can convert to any unseen speakers. FreeVC follows the typical VC paradigm to separate speaker identity
from content and replace the speaker information to perform the conversion. To encourage better separation, FreeVC adopts an information bottleneck and data augmentation to improve the purity of extracted content information. However, learning disentangled representations is still challenging \cite{zhao2020voice} \cite{9413403}. kNN-VC follows another classical paradigm called concatenative VC \cite{jin2016cute}. The concept behind concatenative conversion involves seamlessly joining segments of target speech that correspond to the source content. In kNN-VC, the segments are representations extracted from a self-supervised model WavLM \cite{chen2022wavlm}. As the output is completely constructed from target speech, kNN-VC ensures good speaker similarity. 

However, when the target speech is short, kNN-VC degrades significantly as there are not enough target speech representations to cover the source content. One solution to this problem is to expand the target representations set with synthesized samples conditioned on the initially small set. Several machine learning models for set data have been proposed recently that are based on permutation invariant and equivariant layers \cite{zaheer2017deep}, flow models \cite{bender2020exchangeable}, neural ODE \cite{li2020exchangeable}, kernel methods \cite{shan2022transparent,baskaran2022distribution}, attention mechanism \cite{lee2019set, shan2023nrtsi}, and set expansion methods \cite{zhang2020empower,li2021partially,yan2020end,zaheer2017deep}. However, existing set expansion methods either only focus on generating low-dimensional set data such as 3D point clouds, or directly selecting samples from a candidate pool rather than generating new samples. While in our work we need to generate speech representations that are 1024-dimensional.

Our method to synthesize new speech data to improve speech tasks is also related to existing works. For instance, \cite{hu2022synt++} and \cite{Fazel2021} use synthetic speech to improve ASR performance. However, obtaining high-quality and temporally coherent synthetic speech is challenging. Instead, in our work, we only need to synthesize individual frame-level speech representations that are unordered and do not necessarily have to be coherent, a task that is much easier. 

\begin{figure}[t!]
\centering
\includegraphics[width=0.99\columnwidth]{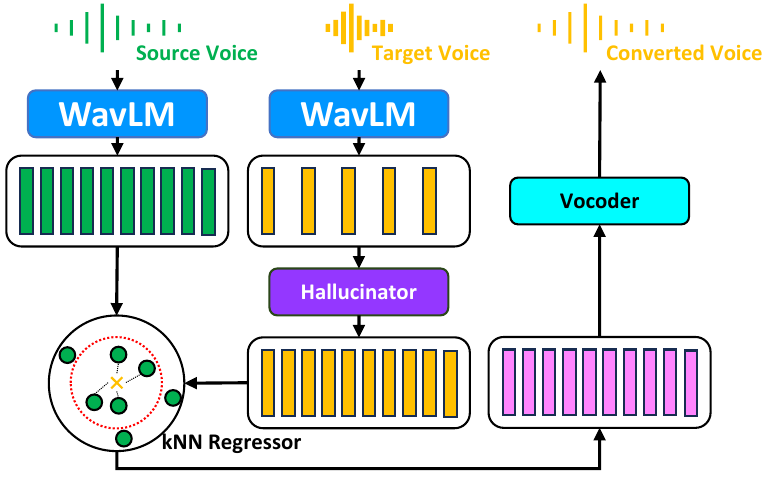}
\caption{The VC pipeline of our method. A pre-trained WavLM model \cite{chen2022wavlm} extracts the source representation sequence (green) from the source voice and the target representation set (yellow) from the target voice respectively. Then, the target set is expanded by our hallucinator. Afterward, every source representation is replaced by its neighbors in the expanded target set, resulting in the converted sequence (pink). Finally, a pre-trained vocoder transforms the converted sequence to voice.}
\label{fig:flow}
\end{figure}

\section{Methods}
\label{sec:method}
\subsection{VC pipeline}
Our novel pipeline is as follows. We adopt the neighbor-based VC method proposed in kNN-VC \cite{baas2023voice}. The overall VC pipeline of \textit{Phoneme Hallucinator} is shown in Fig. \ref{fig:flow}.  Given a source voice and a short target voice, we first use a pre-trained WavLM \cite{chen2022wavlm} model to extract a sequence of frame-level source voice representations $\mathbf{X}^s = (\mathbf{x}_1^s, \mathbf{x}_2^s, ..., \mathbf{x}_{N_s}^s)$ and a set of frame-level target voice representations $\mathbf{X}^t = \{\mathbf{x}_1^t, \mathbf{x}_2^t, ..., \mathbf{x}_{N_t}^t\}$, where every $\mathbf{x} \in \mathbb{R}^d$ denotes a representation of a frame. The sequence length $N_s$ and the set cardinality $N_t$ are proportional to the duration of the source and the target voice. Note that we regard target representations as a set because keeping their temporal order is unnecessary for the subsequent neighbor-based conversion. If the target set is small, we propose to utilize a hallucinator to generate an extra set of features $\mathbf{X}^e = \{\mathbf{x}_1^e, \mathbf{x}_2^e, ..., \mathbf{x}_{N_e}^e\}$ with diversified phonemes belonging to the same target speaker and combine them with the target set, resulting in an expanded target set $\mathbf{X}^{t,e} = \mathbf{X}^t \cup \mathbf{X}^e$. Then, every source feature in $\mathbf{X}^s$ is replaced by an average of its $k$-nearest neighbors in $\mathbf{X}^{t,e}$, resulting in a converted feature sequence $\mathbf{X}^{c} = (\mathbf{x}_1^c, \mathbf{x}_2^c, ..., \mathbf{x}_{N_s}^c)$. Finally, a HiFi-GAN vocoder \cite{kong2020hifi} is adopted to translate the converted features to voice. WavLM  is a self-supervised model that jointly learns masked speech prediction and denoising. After training, it can solve full-stack downstream speech tasks such as speaker verification, speech separation, and  speech recognition. In kNN-VC, it is found that a specific layer of WavLM is capable of  mapping frames of the same phoneme closer to each other than to different phonemes, which ensures the content of the output voice is the same as the source while only changing speaker characteristics. Note that the original kNN-VC does not contain a hallucinator, which leads to deteriorated intelligibility when the target voice is short as discussed in Sec. \ref{sec:intro} and shown in Fig. \ref{fig1}.

\subsection{Hallucinator} 
\label{sec:hallucinator}
The hallucinator is a key component to our VC method. Conditioned on a small set of WavLM representations $\mathbf{X}^t$ from a speaker, it is a generative model that estimates the conditional likelihood $p(\mathbf{X}^e\ |\ \mathbf{X}^t)$ and is capable of sampling an infinite number of WavLM representations $\mathbf{x}^e \sim p(\mathbf{X}^e\ |\ \mathbf{X}^t)$ with diversified phonemes approximately from the same speaker. To build such a model, we adopt De Finetti's theorem \cite{li2021partially,edwards2017towards,qi2017pointnet} to decompose the conditional distribution $p(\mathbf{X}^e\ |\ \mathbf{X}^t)$ into 
\begin{equation}
\label{eq:de_finetti}
    p(\mathbf{X}^e\ |\ \mathbf{X}^t) = \int \prod_{i=1}^{N_e} p(\mathbf{x}^e_i\ |\ \theta) p(\theta\ |\ \mathbf{X}^t) d\theta,
\end{equation}
where set elements are conditionally independent given a multivariate latent variable $\theta$ that contains speaker characteristic information. However, directly optimizing \eqref{eq:de_finetti} is intractable due to the high-dimensional integration over $\theta$. Instead, we optimize a variational lower bound of \eqref{eq:de_finetti} as
\begin{align}
\label{eq:variation}
\begin{split}
    \text{log} p (\mathbf{X}^e\ |\ \mathbf{X}^t) \geq \sum_{i=1}^{N_e} \mathbb{E}_{q(\theta |\mathbf{X}^e, \mathbf{X}^t)} \text{log} p(\mathbf{x}^e_i\ |\ \theta)  \\
    - D_{KL}(q(\theta |\mathbf{X}^e, \mathbf{X}^t)\ ||\ p(\theta\ |\ \mathbf{X}^t)).
\end{split}
\end{align}
Here, $q(\theta |\mathbf{X}^e, \mathbf{X}^t)$ and $p(\theta\ |\ \mathbf{X}^t)$ denotes the posterior distribution and the prior distribution respectively, and $p(\mathbf{x}^e_i\ |\ \theta)$ is a conditional likelihood function over a $\mathbb{R}^d$ speech representation space. Unlike $p(\mathbf{x}^e_i\ |\ \theta)$ which is a probabilistic function for a single feature vector, $q(\theta |\mathbf{X}^e, \mathbf{X}^t)$ and $p(\theta\ |\ \mathbf{X}^t)$ requires probabilistic modeling for a set of feature vectors, which is more challenging as a set is a collection of exchangeable elements with arbitrary cardinality. Our model also needs to respect the unordered property of sets, i.e. the probability functions $q(\theta |\mathbf{X}^e, \mathbf{X}^t)$ and $p(\theta\ |\ \mathbf{X}^t)$ should be invariant even if we randomly permute elements in sets. 

\begin{figure*}[t!]  
\centering
\includegraphics[width=1.29\columnwidth]{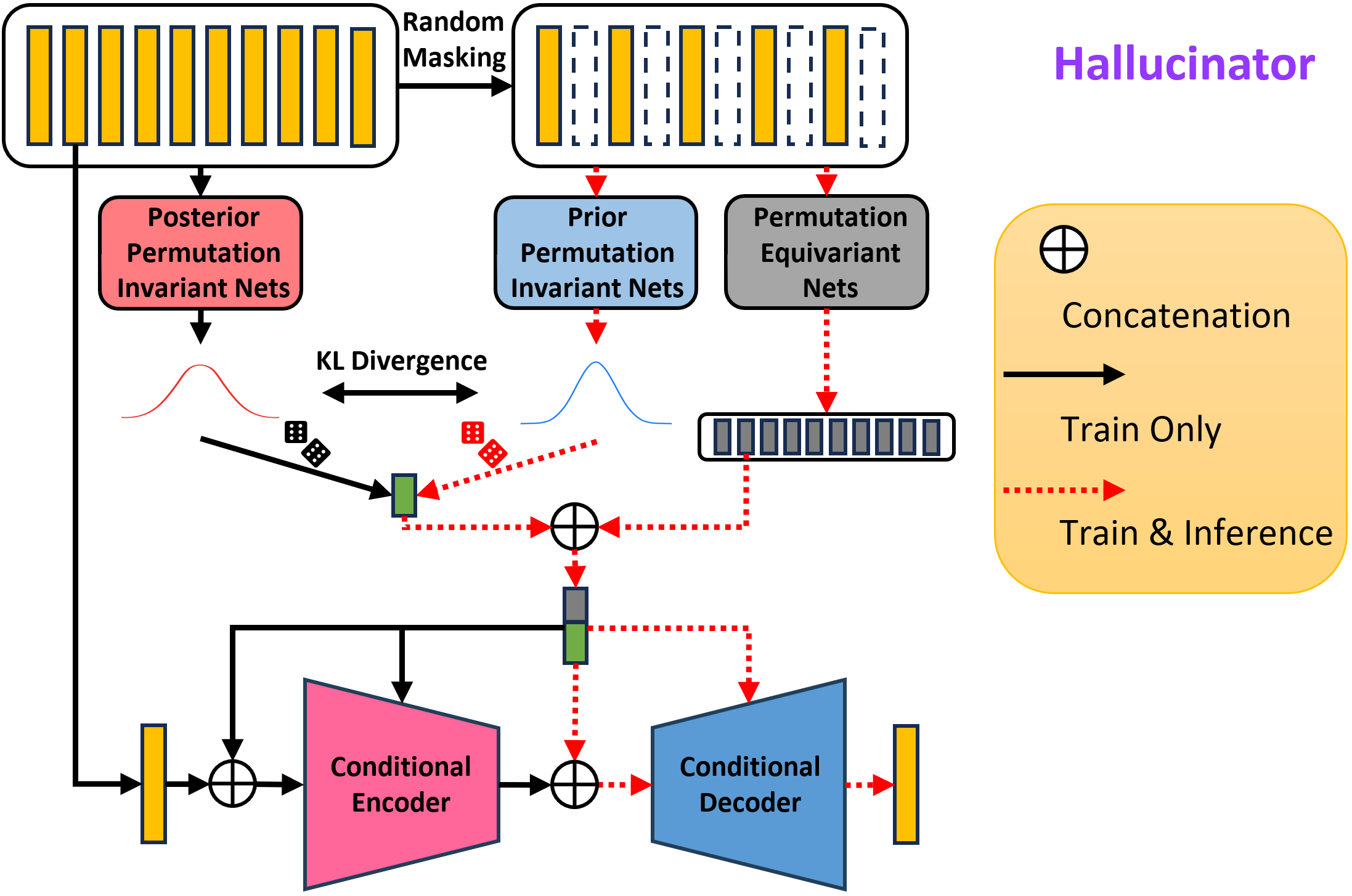}
\caption{The detailed structure of the hallucinator. Posterior Permutation Invariant Nets, Prior Permutation Invariant Nets, and Permutation Equivariant Nets are all implemented by Set Transformer. Conditional Encoder and Conditional Decoder are implemented by multilayer perceptrons (MLP).}
\label{fig:hallucinator}
\end{figure*}

Neural networks tailored for set data have been developed recently, such as Deep Sets \cite{zaheer2017deep}, Set Transformer \cite{lee2019set}, and ExNODE \cite{zaheer2017deep}. Among these methods, Set Transformer uses multi-head self-attention mechanism to model the interactions between set elements and introduces an inducing point method to avoid the quadratic time complexity w.r.t. set cardinality. We adopt Set Transformer to implement $q(\theta |\mathbf{X}^e, \mathbf{X}^t)$ and $p(\theta\ |\ \mathbf{X}^t)$ due to its flexibility and speed. 

For the conditional generative function $p(\mathbf{x}^e_i\ |\ \theta)$, we could use conditional VAE, diffusion models, or normalizing flow models. For diffusion models, they usually require several iterations to obtain high-quality samples, which may slow down the inference speed. For normalizing flow methods, their latent representations usually have the same size as the data, which may significantly increase the computation burden when the dimensionality of data is huge. For these considerations, we implement $p(\mathbf{x}^e_i\ |\ \theta)$ using a conditional VAE decoder model with a fast sampling speed, low computation burden, and losses that can be conveniently optimized.

We leave the discussions about implementation details of the posterior, the prior, and the generative functions to Sec. \ref{sec:implementation}.

\subsection{Training and Inference} 
\label{sec:train_inference}
Among all the components in our pipeline (shown in Fig. \ref{fig:flow}), only WavLM, Hallucinator, and Vocoder are trainable while the kNN regressor doesn't require any training. In this work, we find using a WavLM pre-trained in \cite{chen2022wavlm} and a prematched Vocoder pre-trained in \cite{baas2023voice} without further fine-tuning already achieves good performance and we only need to train the Hallucinator from scratch. 

To train Hallucinator, we need to obtain an extra set of speech representations $\mathbf{X}^e$ conditioned on $\mathbf{X}^t$. Equivalently, we can reformulate this task as obtaining a missing subset of speech representations $\mathbf{X}^e$ based on an observed subset of speech representations $\mathbf{X}^t$. Both $\mathbf{X}^e$ and $\mathbf{X}^t$ are from the same speech utterance representation $\mathbf{X}$ with a fixed cardinality $N$ where $\mathbf{X}^e \cap \mathbf{X}^t = \emptyset$ and $\mathbf{X} = \mathbf{X}^e \cup \mathbf{X}^t$. 

Though speech representations extracted from a continuous utterance are usually temporally correlated, we assume that the missing elements are \textit{Missing Completely at Random (MCAR)}. Our assumption is valid because our task is not to synthesize a sequence of temporally coherent speech representations, but to synthesize a set of unordered representations with diversified phonemes from the same speaker. Therefore, we obtain $\mathbf{X}^e$ and $\mathbf{X}^t$ during training as follows: 
\begin{enumerate}
    \item \textit{uniformly choose the number of missing representations $N_e$ with $1 \leq N_e \leq N$.}
    \item \textit {uniformly select $N_e$ elements from $\mathbf{X}$, resulting in $\mathbf{X}^e$.}
    \item \textit{mask out the elements selected above in $\mathbf{X}$ by zero vectors, resulting in $\mathbf{X}^t$. This step is visualized in the top part of Fig. \ref{fig:hallucinator}. To differentiate selected and un-selected elements, each element in $\mathbf{X}^t$ is concatenated with a binary mask.} 
\end{enumerate}

After obtaining $\mathbf{X}^t$ and $\mathbf{X}^e$, we can train our model by maximizing \eqref{eq:variation}. 
During inference, our model takes $\mathbf{X}^t$ and outputs $\mathbf{X}^e$. Note that we can run as many sampling steps as we want to obtain an infinite number of new samples.

\subsection{Model Implementation} 
\label{sec:implementation}
An illustration of our Hallucinator is given in Fig. \ref{fig:hallucinator}. For both the prior and the posterior in \eqref{eq:variation}, we use Set Transformer to compute a permutation equivariant embedding for each set element and then average them over the set to produce permutation invariant parameters to a multivariate Gaussian distribution. The mean and variance of the Gaussian are derived from the set representation. To further increase the expressiveness of the prior $p(\theta\ |\ \mathbf{X}^t)$ so that it can better match the posterior $q(\theta |\mathbf{X}^e, \mathbf{X}^t)$, the prior is augmented with a normalizing flow model with base distribution defined as a Gaussian conditioned on the set representation. Inspired by POEx \cite{li2021partially}, we also use another Set Transformer to extract a permutation equivariant embedding $\mathbf{g}_i$ for each element in $\mathbf{X}^t$ and concatenate $\mathbf{g}_i$ with $\theta$ to assist the learning of $p(\mathbf{x}^e_i\ |\ \theta)$. Now, $p(\mathbf{x}^e_i\ |\ \theta)$ in \eqref{eq:variation} is rewritten as $p(\mathbf{x}^e_i\ |\ \theta, \mathbf{g}_i)$. We implement it using a conditional VAE as discussed in Sec. \ref{sec:hallucinator}. We derive a variational lower bound of $p(\mathbf{x}^e_i\ |\ \theta, \mathbf{g}_i)$ as 
\begin{align}
\label{eq:generative}
\begin{split}
    \text{log} p(\mathbf{x}^e_i\ |\ \theta, \mathbf{g}_i) \geq \mathbb{E}_{q(z|\mathbf{x}^e_i, \theta, \mathbf{g}_i)} \text{log} p(\mathbf{x}^e_i\ |\ z, \theta, \mathbf{g}_i) \\
    - D_{KL}(q(z\ |\ \mathbf{x}^e_i, \theta, \mathbf{g}_i)\ ||\ p(z\ |\ \theta, \mathbf{g}_i)),
\end{split}
\end{align}
where $p(z\ |\ \theta, \mathbf{g}_i)$ and $q(z\ |\ \mathbf{x}^e_i, \theta, \mathbf{g}_i)$ respectively denote conditional prior and posterior distributions that are multivariate Gaussian. $p(\mathbf{x}^e_i\ |\ z, \theta, \mathbf{g}_i)$ is a likelihood function from a multivariate Gaussian in a $\mathbb{R}^d$ space, and $z$ is a latent code containing speech information of $\mathbf{x}^e_i$, such as its phoneme content, pitch, and volume. To condition the prior on $\theta$ and $\mathbf{g}_i$, $z$ is concatenated with $\theta$ and $\mathbf{g}_i$. $p(\mathbf{x}^e_i\ |\ z, \theta, \mathbf{g}_i)$ and $q(z\ |\ \mathbf{x}^e_i, \theta, \mathbf{g}_i)$ are implemented using multi-layer perceptrons (MLP) decoder and encoder respectively. To condition the encoder on $\theta$ and $\mathbf{g}_i$, we concatenate them with $\mathbf{x}^e_i$. To further encourage the encoder and the decoder to exploit the speaker information contained in $\theta$ and $\mathbf{g}_i$ to reconstruct $\mathbf{x}^e_i$, we modulate each layer of MLP as follows:
\begin{equation}
\label{eq:cond}
c_l \leftarrow \text{Sigmoid}(W_{c,l}^T \theta + W_{g,l}^T \mathbf{g}_l + b_{c,l}),
\end{equation}
\begin{equation}
h_l \leftarrow \text{LeakyReLU}(W_{l}^T h_{l-1} + b_{l}),
\end{equation}
\begin{equation}
h_l \leftarrow h_l \odot c_l,
\end{equation}
where $h_l$ denotes the feature at layer $l$ and it is modulated by element-wise multiplication with $c_l$. A sigmoid function is used in \eqref{eq:cond} to improve training stability. As will be shown in Sec. \ref{sec:ablation}, we find both concatenation and modulation are necessary to better condition on set representations $\theta$ and $\mathbf{g}_i$ to improve speaker similarities.

\begin{table*}[t!]
\centering
\begin{tabular}[t]{lccccc}
\toprule
& WER$\downarrow$ & CER$\downarrow$ & EER$\uparrow$ & MOS$\uparrow$ & SIM$\uparrow$\\
\midrule
Oracle (Real speech from test split) & 3.47\% & 1.42\% & --- & 4.27$\pm$0.10 & 3.25$\pm$0.09 \\
\midrule
VQMIVC \cite{wang2021vqmivc} & 41.51\% & 23.66\% & 11.84\% & 2.82$\pm$0.11 & 2.10$\pm$0.13\\
YourTTS \cite{casanova2022yourtts} & 8.65\% & 3.36\% & 38.23\% & 3.41$\pm$0.11 & 2.71$\pm$0.11 \\
kNN-VC \cite{baas2023voice} & 45.92\% & 27.55\% & 19.19\% & 2.34$\pm$0.17 & 2.48$\pm$0.12 \\
FreeVC \cite{li2023freevc} & 5.40\% & 2.27\% & 35.63\% & \textbf{4.05}$\pm$\textbf{0.09} & 2.62$\pm$0.11 \\
\midrule
\textit{Phoneme Hallucinator (Ours)} & \textbf{5.10\%} & \textbf{2.02\%} & \textbf{44.62\%} & \textbf{4.02}$\pm$\textbf{0.08} & \textbf{2.98}$\pm$\textbf{0.12} \\

\bottomrule
\end{tabular}
\caption{One-shot VC performance given only 3 seconds of target speech. Standard deviations are shown for MOS and SIM.}
\label{tab:result}
\end{table*}

\section{Experiments}
In this section, we attempt to answer the following research questions:
\begin{enumerate}
    \item Can our model outperform existing methods w.r.t. both \textit{intelligibility} and \textit{speaker similarity} when the target speech is short?
    \item Can our model hallucinate representations that are \textit{diversified} enough to cover a wide range of phonemes, while at the same time \textit{faithfully preserving} the target speaker characteristics? 
    \item How does each component of our model contribute to the performance?
\end{enumerate}

To answer these questions, we train our model on LibriSpeech \textit{train-clean-100} split and evaluate on LibriSpeech \textit{test-clean} split, which contains 40 speakers unseen during training and each speaker contains approximately 8 minutes of 16 kHz English speech \cite{panayotov2015librispeech}. Given the diversity of speakers, this experiment setup can provide a thorough evaluation of different any-to-any VC methods. 

\subsection{Experimentation Details} We use the pre-trained WavLM-large encoder \cite{chen2022wavlm} to extract frame-level speech representations where a single 
1024-dimensional vector is produced for every 20ms of 16 kHz audio. Following kNN-VC \cite{baas2023voice}, we use the feature from the 6th layer of WavLM, which contains both speaker information and phoneme information. For the kNN regressor, we set $k$ equal to 4. We adopt the HiFi-GAN Vocoder \cite{kong2020hifi} pre-trained in kNN-VC where it is trained on prematched 6th layer features \cite{baas2023voice}.

To train our model, we only use utterances longer than 4 seconds, which ensures that we can extract a representation set with a cardinality larger than 200. We randomly select a subset of 200 represents to constitute $\mathbf{X}$, which will be divided into the missing subset $\mathbf{X}^e$ and the observed subset $\mathbf{X}^t$ to train our model as discussed in Sec. \ref{sec:train_inference}. During every inference step, if the cardinality $N_t$ of the target set is greater than 100, we uniformly select a subset with a cardinality of 100 to form $\mathbf{X}_t$. To facilitate learning, we divide the representations by 10 to normalize them to have a standard deviation of around 1. All Set Transformers contain 4 multi-head attention blocks with a hidden size of 256 and 16 induction points. The sizes of $\theta$ and $\mathbf{g}_i$ are 256. The size of $z$ is 256 and both the decoder and the encoder in \eqref{eq:generative} is a 4-layer MLP with a hidden size of 512. We do not use any normalization layer for both MLP and Set Transformer. Our Hallucinator is implemented by TensorFlow and trained by Adam optimizer with a learning rate of 0.0001 and a batch size of 50 for 250 epochs on an NVIDIA RTX 4090 GPU. We train our model by maximizing the log likelihood function defined in \eqref{eq:variation}. Training takes 6.5 hours. The total number of trainable parameters is 19.03M for our Hallucinator. LibriSpeech \textit{dev-clean} split is used as the validation set. 

\subsection{Baselines} We compare the \textit{Phoneme Hallucinator} to text-free any-to-any VC methods, such as kNN-VC\footnote{\url{https://github.com/bshall/knn-vc}} \cite{baas2023voice}, FreeVC\footnote{\url{https://github.com/OlaWod/FreeVC}} \cite{li2023freevc}, VQMIVC\footnote{\url{https://github.com/Wendison/VQMIVC}} \cite{wang2021vqmivc} and YourTTS\footnote{\url{https://github.com/Edresson/YourTTS}} (in text-free conversion mode) \cite{casanova2022yourtts} using their official implementations. Among these methods, FreeVC, YourTTS, and VQMIVC use speaker encoders to extract target speaker information for conversion, while kNN-VC adopts the classical concatenative VC approach. FreeVC, YourTTS and VQMIVC respectively use data augmentation, text transcriptions during training, and vector quantization to separate content and speaker information.

\subsection{Evaluation Metrics} 
\subsubsection{Objective Metrics} To report objective metrics, we use all 40 speakers from LibriSpeech \textit{test-clean} split and randomly select 5 utterances for each speaker. For each utterance, we convert it to all the utterances of other speakers, resulting in a total of 39,000 (40$\times$5$\times$39$\times$5) converted speech utterances.

We follow kNN-VC to use Equal Error Rate (EER) to objectively measure speaker similarity. With a trained speaker verification system\footnote{\url{https://huggingface.co/speechbrain/spkrec-xvect-voxceleb}} that computes an x-vector \cite{snyder2018x} of speech to represent speaker identity, we compute cosine similarities between a pair of x-vectors to measure speech similarity. For all converted utterances, we compute their similarity scores to their corresponding target utterances. Then we compute an equal number of ground-truth similarity scores, which are computed from pairs of two randomly sampled utterances from the same target speaker. Finally, an EER is computed by combining the above two sets of scores, assigning a label of 1 to the ground-truth pairs and 0 to pairs containing converted speech. A higher EER means better speaker similarity and the maximum EER is 50\%, meaning that the converted speech is indistinguishable from the real speech from the target speaker.
\begin{figure*}[h!]
\centering
\begin{subfigure}{0.33\textwidth}
    \includegraphics[width=\textwidth, trim={2cm 2cm 2cm 2cm}, clip]{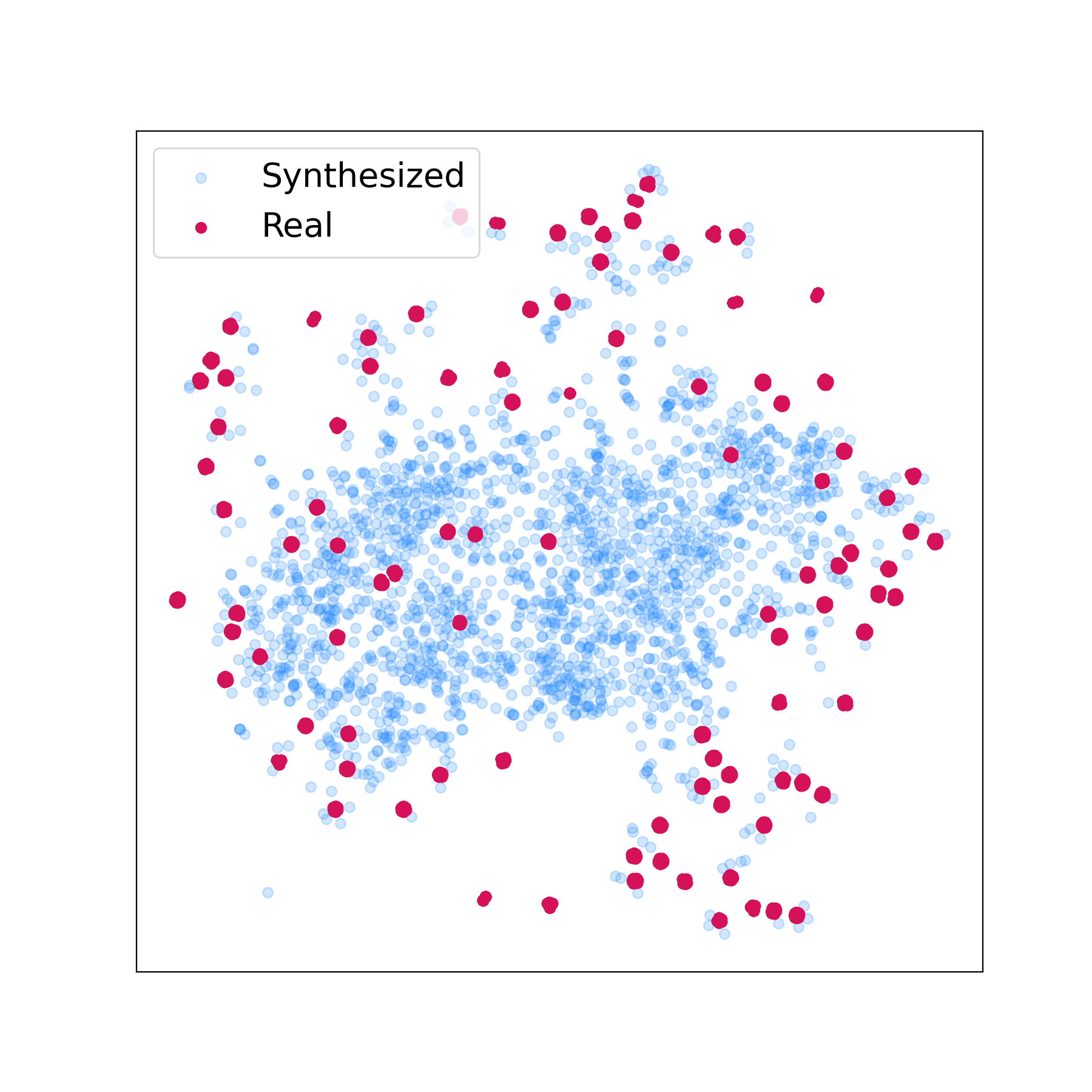}
\end{subfigure}
\begin{subfigure}{0.33\textwidth}
    \includegraphics[width=\textwidth, trim={2cm 2cm 2cm 2cm}, clip]{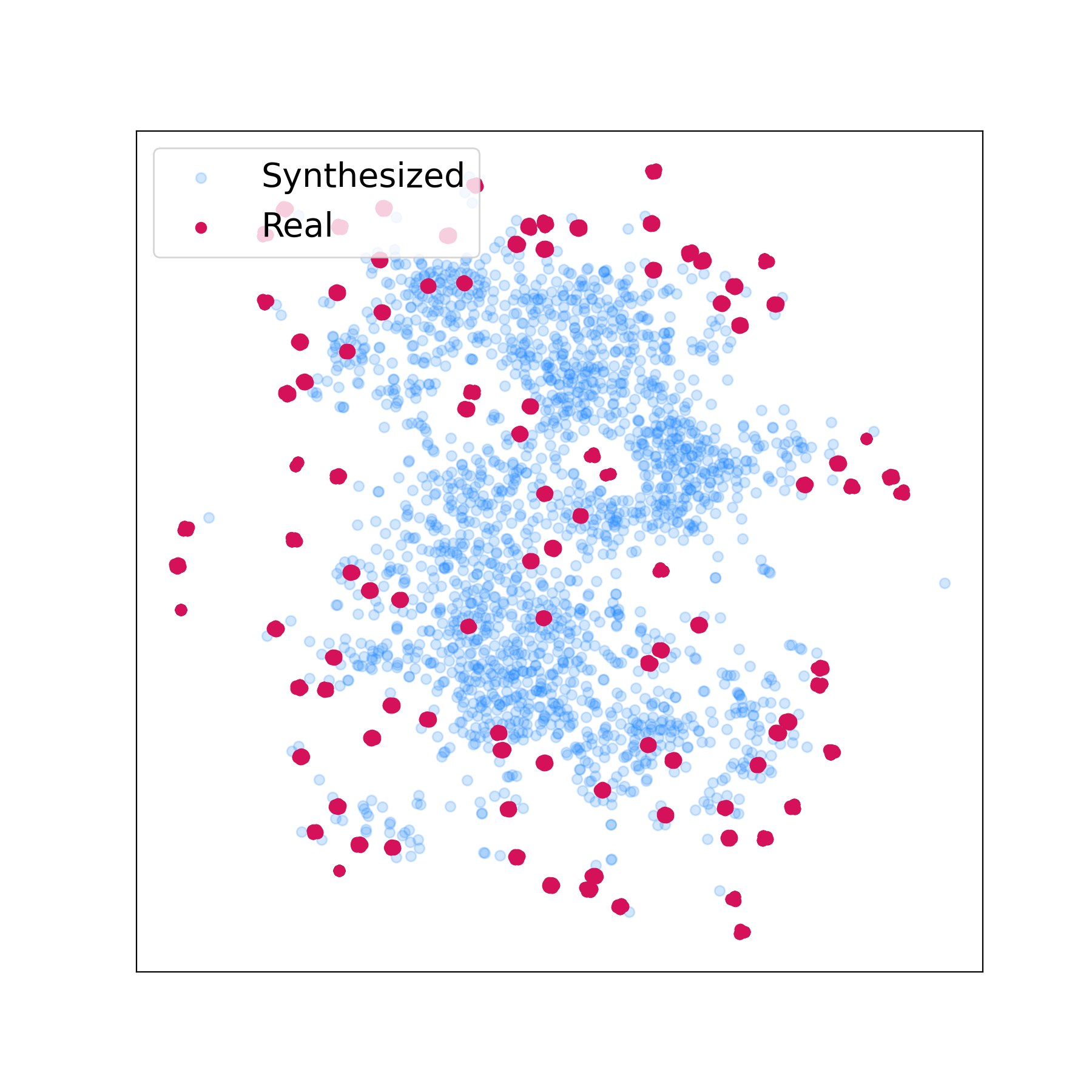}
\end{subfigure}
\begin{subfigure}{0.33\textwidth}
    \includegraphics[width=\textwidth, trim={2cm 2cm 2cm 2cm}, clip]{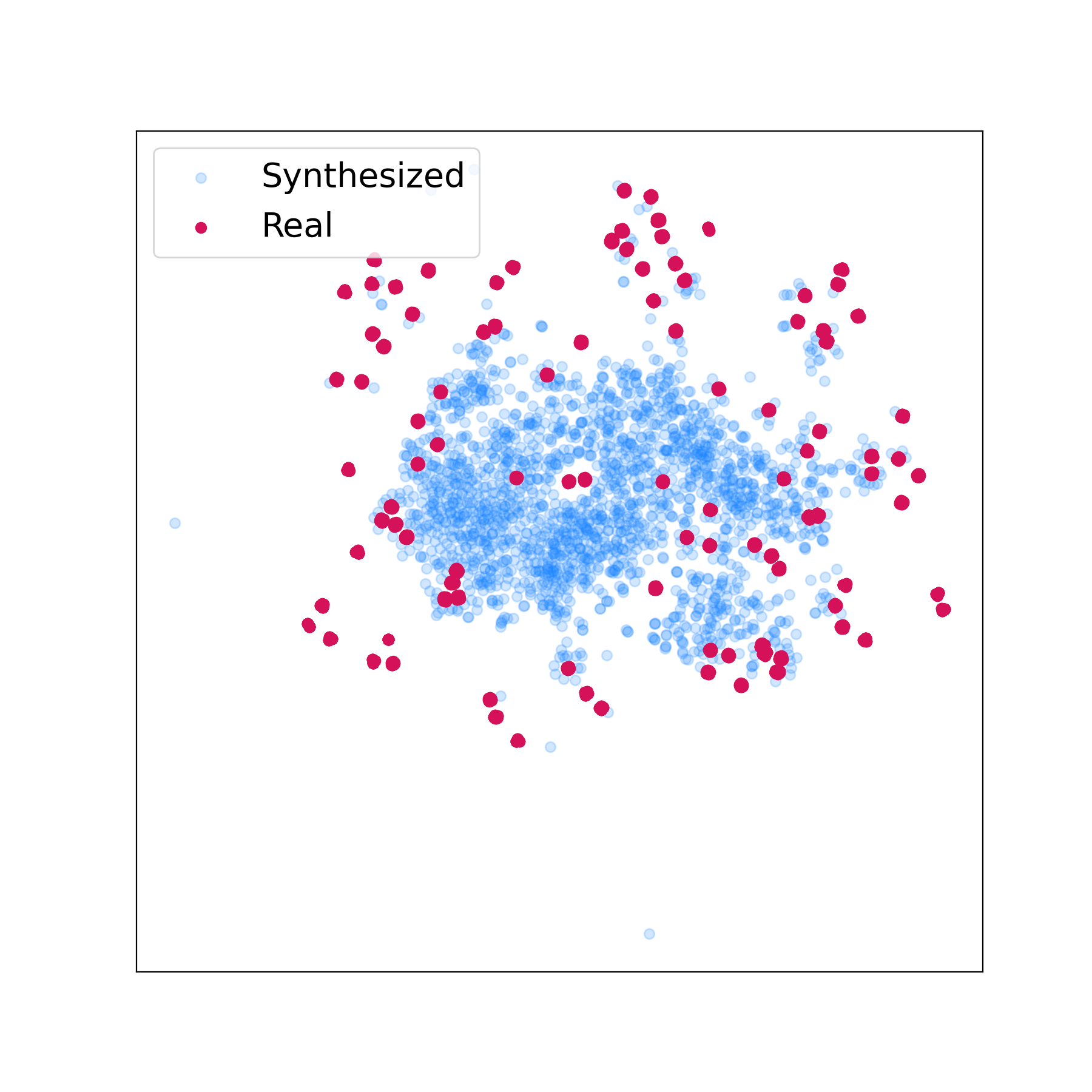}
\end{subfigure}
  \caption{T-SNE visualization of three randomly chosen expanded sets. In each subplot, there are 100 given target speech representations (red) extracted from a 2-second utterance from a speaker in LibriSpeech \textit{test-clean} split, which equals 2 seconds of speech. Conditioned on the given representations, our model hallucinates 2,000 new representations (blue).}
  \label{fig:tsne}
\end{figure*}

To objectively measure intelligibility, we use an ASR system\footnote{\url{https://github.com/openai/whisper}} \cite{radford2023robust} to transcribe the converted speech and compute the word/character error rate (WER/CER). Lower error rates indicate better intelligibility.

\subsubsection{Subjective Metrics} We use Amazon Mechanical Turk to subjectively measure naturalness and speaker similarity. For naturalness, we adopt mean opinion scores (MOS) in the range 1–5, where 1 is the lowest perceived quality, and 5 is the highest perceived quality. We randomly sample 50 utterances for each method. For speaker similarity, given a pair of utterances, we ask listeners to judge speaker similarity in the range 1-4. We randomly sample 20 source utterances from the LibriSpeech \textit{test-clean} split and convert each of them to 4 target speakers. We report the average of this speaker similarity metric and denote it as SIM. 

\subsection{One-shot VC results} To evaluate one-shot VC performance, we simulate the scenario where the target speech is short by only using the first 3 seconds of each utterance when it is used as the target speaker's speech. Note that our one-shot VC task is much more challenging than the VC task considered in kNN-VC where all the available target data (around 8 minutes per speaker) is used for conversion. Besides, our one-shot VC task fit real-world application scenario better as users usually can not provide an abundant amount of target speaker data. 

\subsubsection{Answer to Intelligibility and Speaker Similarity}

For a short given target speech, we use our Halluscinator to sample extra 30,000 new representations, which is approximately equal to 10 minutes of extra speech. We report the performance of all methods in Table \ref{tab:result}. We find that \textit{Phoneme Hallucinator} simultaneously achieves SOTA performance on all metrics that measure intelligibility, speaker similarity and naturalness. This answer the first question at the beginning of this section. Compared to our method, kNN-VC degrades significantly with such a short target speech and its converted speech is almost unintelligible as indicated by its high WER. FreeVC is the best-performing baseline and it achieves a comparable MOS score to our method. However, it is not as intelligible as our method and its speaker similarity is worse. \textit{We encourage the reader to listen to the converted audio samples of all the methods in supplemental materials.}

\subsubsection{Answer to Hallucination Diversity and Faithfulness}

To answer the second question at the beginning of this section, we report the intelligibility and speaker similarity of \textit{Phoneme Hallucinator} with varying target voice duration from 1 second to 10 seconds in Fig. \ref{fig1}. It can be seen that our method consistently maintains a low WER of around 5\%, which indicates that our hallucinated features are diversified enough to cover the source speech features. Compared to kNN-VC, which only uses realistic target features rather than synthetic features for conversion, our method achieves a comparable and sometimes even better speaker similarity. This indicates that our hallucinated features also faithfully preserve the target speaker's characteristics.

To further show the diversity of hallucinated features, we visualize them by t-SNE \cite{van2008visualizing} in Fig. \ref{fig:tsne}. Given only 100 target speech features (shown in red) extracted from a 2-second utterance from a speaker in LibriSpeech \textit{test-clean} split, we use our model to hallucinate 2,000 new features and find that the hallucinated features (shown in blue) effectively cover the large blank area between the given features. Another interesting observation is that our model tends to avoid synthesizing duplicated features that already exist in the given target set. This phenomenon is probably caused by the way our model is trained, which requires that $\mathbf{X}^e \cap \mathbf{X}^t = \emptyset$ according to Sec. \ref{sec:train_inference}.

\subsubsection{Inference Time} With an NVIDIA RTX 4090 GPU and an Intel i9-13900F CPU, our hallucinator only takes 1.48 seconds to hallucinate 30,000 new representations, which is approximately equivalent to 10 minutes of speech. Note that given a single target speech and several different source speeches, we only need to hallucinate once and re-use the hallucination results. The inference times of other components of our model, namely WavLM, kNN regressor and vocoder, are the same as those in kNN-VC.

\subsection{Ablation Studies}

\begin{table}[t]
\centering
\begin{tabular}[t]{lcccccc}
\toprule
\#Samples& 5k & 10k & 15k & 20k & 30k & 50k \\
\midrule
WER$\downarrow$ (\%) & 6.14 & 5.62 & 5.35 & 5.24 & 5.10 & \textbf{4.84} \\
EER$\uparrow$ (\%) & \textbf{49.1} & 47.2 & 46.2 & 44.4 & 44.6 & 43.3 \\
\bottomrule
\end{tabular}
\caption{The influence of the number of hallucinated features on Word Error Rate (WER) and Equal Error Rate (EER), which respectively measure intelligibility and speaker similarity.}
\label{tab:num_samples}
\end{table}

\subsubsection{Number of Hallucinated Samples}
\label{sec:ablation}
We can freely choose how many representations to synthesize to control the target set cardinality. In Table \ref{tab:num_samples}, we find increasing the cardinality leads to an increase in intelligibility but a slight decrease in speaker similarity. The cause of this phenomenon is that with more hallucinated target representations, each source representation can find closer neighbors from the expanded target set, which biases the converted speech towards the source speech. We emphasize that even if the target set only consists of realistic representations, this phenomenon still exists as shown by Figure 2 in \cite{baas2023voice}. In fact, for a large target set from a single speaker without any hallucination, there are still variations in tone, emotion and pitch within the set, which will bias the converted speech to the source. Therefore, the cause of this phenomenon is the neighbor-based pipeline rather than our hallucinated representations. For a specific application scenario, we can tune the cardinality to find an optimal trade-off between speaker similarity and intelligibility.

\subsubsection{Model Architecture} Here we investigate the contributions of the permutation equivariant embedding $\mathbf{g}_i$ (PEQ), the concatenation mechanism (CAT), and the modulation mechanism (MOD) introduced in Sec. \ref{sec:implementation}. With all model variants hallucinating 30,000 representations, we report their experiment results in Table \ref{tab:ablation}. V1 is our proposed method with all the components enabled. Comparing V3 to V1 in Table \ref{tab:ablation}, we can see CAT is necessary to achieve good speaker similarity. Comparing V4 to V1, we find that MOD is beneficial to improve speaker similarity. Comparing V2 to V1, we find removing PEQ is helpful to improve speaker similarity, but at the cost of reducing intelligibility. Also, we find that removing PEQ harms training stability as the model will get stuck in local minima for a while at the early training stage. In cases where speaker similarity is more important, we can remove PEQ and increase the number of hallucinated features to improve intelligibility. We find that our model w/o PEQ with 100k hallucinated features achieves a WER of 5.17\% and an EER of 47.20\%. Note that we do not consider the case where both CAT and MOD are removed, which degrades our conditional VAE to a vanilla VAE and completely ignores the given target set to achieve good speaker similarity.

\begin{table}[t]
\centering
\begin{tabular}[t]{cccccc}
\toprule
Variant & PEQ & CAT & MOD & WER$\downarrow$ & EER$\uparrow$  \\
\midrule 
V1 & $\checkmark$ & $\checkmark$ & $\checkmark$ & \textbf{5.10\%} & 44.62\%  \\
V2 & & $\checkmark$ & $\checkmark$ & 5.50\% & \textbf{48.10\%} \\
V3 & $\checkmark$ & & $\checkmark$ & 5.45\% & 17.59\%   \\
V4 & $\checkmark$ & $\checkmark$ &  & 5.59\% &  39.96\%  \\
V5 & & $\checkmark$ & & 5.37\% &  36.80\% \\
V6 &  &  & $\checkmark$ & 5.90\% & 18.60\%  \\

\bottomrule
\end{tabular}
\caption{Ablation studies of model architecture. PEQ, CAT, and MOD denote permutation equivariant embedding, concatenation, and modulation respectively.}
\label{tab:ablation}
\end{table}

\section{Conclusion, Discussion and Future Works}

In this paper, we propose \textit{Phoneme Hallucinator}, a novel conditional set generative model that can hallucinate speech representations based on a small set of target speech representations. Therefore, our model effectively addresses the intelligibility deterioration issue in kNN-VC when the target speech is short. At the same time, as our hallucinated representations also faithfully preserve the target speaker's characteristics, the \textit{Phoneme Hallucinator} inherits the advantage of kNN-VC where a high speaker similarity is achieved. Taken together, we model achieves SOTA one-shot VC performance regarding speaker similarity, intelligibility, and naturalness for both subjective and objective metrics. \textit{Phoneme Hallucinator} is fast to train and inference. We also perform extensive ablation studies to investigate the contribution of each design in our model and show evidences that our model can hallucinate diversified representations to cover the source representations. Audio samples are provided in supplemental materials and we strongly encourage readers to listen to them.

Though our hallucinator is only trained on English data, we find it is capable of performing high-quality cross-lingual VC in other languages as well (audio examples are provided), such as Chinese to Germany, Spanish to Chinese, and Germany to Spanish. This is feasible because different languages share a similar set of speech representations. 

There are several future work directions. Currently, we directly use the HiFi-GAN vocoder trained in kNN-VC, which is trained only on real representations rather than our hallucinated representations. As our hallucinated representations are realistic enough, this design already achieves SOTA performance. We believe that the performance could be further improved if we train a vocoder on our hallucinated representations. Another direction is to improve the adaptation capacity \cite{finn2017model, ha2017hypernetworks} of our hallucinator so that it can be quickly adapted to synthesize representations from unseen distributions, e.g. dog barking or whispers. Also, in the future, conditional diffusion models. Though our conditional VAE already works well, it can also be replaced by other generative models, such as conditional diffusion models with fast sampling speed \cite{song2023consistency}, to achieve even better sample quality. 

\section{Acknowledgments} This research was partly funded by NSF grants IIS2133595, DMS2324394, and by NIH grants 1R01AA02687901A1, 1OT2OD032581-02-321.

\section{Ethics statement} Using AI for voice conversion raises ethical issues involving privacy, consent, potential misuse, bias, and accessibility. It's essential to ensure voices are not manipulated without permission or used deceitfully. AI algorithms must also be bias-free to avoid unfair results. Additionally, the unequal access to this technology necessitates guidelines to ensure ethical and beneficial use.

\bibliography{aaai24}

\end{document}